\documentclass[conference,letterpaper,10pt]{IEEEtran}

\usepackage[latin1]{inputenc}
\usepackage{times,amsmath}
\usepackage{amsfonts}
\usepackage{pstool}
\usepackage{subfigure}
\usepackage{multirow}
\usepackage{enumerate}
\usepackage{graphicx}
\usepackage{MnSymbol}
\usepackage{stfloats}
\usepackage[table]{xcolor}
\usepackage[square, comma, sort&compress, numbers]{natbib}
\usepackage{nohyperref}
\usepackage{algorithm,algorithmic}
\usepackage{bigints}
\usepackage{amsmath}

\newcounter{tempEquationCounter}
\newcounter{thisEquationNumber}

\makeatletter
\newcommand{\vast}{\bBigg@{4}}
\newcommand{\Vast}{\bBigg@{5}}
\makeatother

\graphicspath{ {Figures/} }

\begin{document}

\title{Protocol Stack Perspective For Low Latency and Massive Connectivity in Future Cellular Networks}

\author{
\IEEEauthorblockN{Syed Waqas Haider Shah\IEEEauthorrefmark{1}, Adnan Noor Mian\IEEEauthorrefmark{1}\IEEEauthorrefmark{2}, Shahid Mumtaz\IEEEauthorrefmark{6}, M. Wen\IEEEauthorrefmark{5}, T. Hong\IEEEauthorrefmark{3}, M. Kadoch\IEEEauthorrefmark{4}}
\IEEEauthorblockA{\IEEEauthorrefmark{1}Department of Electrical Engineering, Information Technology University, Lahore (54000), Pakistan \\}
\IEEEauthorblockA{\IEEEauthorrefmark{2}Computer Laboratory, 15 JJ Thomson Avenue, Cambridge, University of Cambridge, CB3 0FD, UK \\}
\IEEEauthorblockA{\IEEEauthorrefmark{6}Instituto de Telecommuninicoes, DETI, Universidade de Aveiro, Aveiro (4554), Portugal\\}
\IEEEauthorblockA{\IEEEauthorrefmark{5}School of Electronic Engineering, South China University of Technology, Guangzhou 510640, China}
\IEEEauthorblockA{\IEEEauthorrefmark{3}School of Electronic and Information Engineering, Beihang University, Beijing, China\\}
\IEEEauthorblockA{\IEEEauthorrefmark{4}Department of Electrical Engineering, Ecole de Technologie Superieure, Montreal, Quebec, Canada}
(waqas.haider, adnan.noor)@itu.edu.pk, smumtaz@av.it.pt, eemwwen@scut.edu.cn, \\ hongtao@buaa.edu.cn, michel.kadoch@etsmtl.ca}
\maketitle
\begin{abstract}
With the emergence of Internet-of-Things (IoT) and ever-increasing demand for the newly connected devices, there is a need for more effective storage and processing paradigms to cope with the data generated from these devices. In this study, we have discussed different paradigms for data processing and storage including Cloud, Fog, and Edge computing models and their suitability in integrating with the IoT. Moreover, a detailed discussion on low latency and massive connectivity requirements of future cellular networks in accordance with machine-type communication (MTC) is also presented. Furthermore, the need to bring IoT devices to Internet connectivity and a standardized protocol stack to regulate the data transmission between these devices is also addressed, while keeping in view the resource-constraint nature of IoT devices.
\end{abstract}
\section{Introduction}
It has been more than a decade since the term Internet-of-Things (IoT) was coined. However, despite the momentous and substantial amount of exertions, the notion of IoT seems unaccomplished. The primary reason behind this inadequacy is the non-existence of a centralized and standardized set of protocols to completely govern the IoT platforms. Although, several research groups are striving to bring-forth more optimized versions of the existing Internet protocols, the trend among the developers' community is focused on utilizing existing sub-optimal models to deal with the IoT.

Another important issue that needs to be addressed is that as IoT reaches ubiquity, the massive amount of data generated as a result, would need to be processed and stored efficiently, in order to make a profound sense out of it. Since the inception of IoT, an unprecedented and constant escalation in the generated data is being witnessed. International Data Corporation (IDC) forecasts that by 2025, the total generated data traffic would reach 163 zettabytes, while IoT would account for 95\% of the real-time data \cite{reinsel2017data}. This data would originate from various forms of wireless sensor networks (WSN), vehicular networks, healthcare, and home appliances and can provide valuable insights, underlying patterns, and correlation between different variables. If there is an efficient way to store and analyze these huge volumes of data, organizations can utilize it to enhance their business intelligence and market analysis skills. Cloud computing can be one of the most viable options for handling huge volumes of IoT data. The cloud can also offer financial relief to the organizations as it saves the costs incurred in setting up large infrastructures including expensive software licenses, hardware, networking along with IT support, and maintenance.

Previously, various protocol stacks have been proposed in the literature. The authors in \cite{palattella2013standardized} have proposed a protocol stack, which is mainly focused on applications operating in unlicensed frequency bands. Moreover, while proposing the protocol stack, the authors did not address the issues related to big-data management and data analytics. With the emergence of cognitive radio (CR) communication, it has become one of the key enabling technologies for IoT networks. The authors in \cite{aijaz2015cognitive} have proposed a protocol stack for machine-to-machine (M2M) communication using CR as a physical layer technology. Since, CR is a newer technology that employs opportunistic ways of communication, it does not provide quality-of-service (QoS) therefore, it is not an efficient approach for various critical IoT (cIoT) applications. {\it However, to the best of the authors' knowledge, there is no such protocol stack in the literature which provides efficient data management and data analytics as well as QoS provisioned communication using cellular network at its lower layers}.

This work has the following contributions: i) A detailed discussion on the feasibility of different data processing paradigms (cloud, fog, and edge) for smart IoT is presented; ii) Introduction of data analytics and semantic layer in the existing protocol stack for IoT as shown in Fig. 2; iii) Identification of various issues in IoT protocols for different layers which may bound low latency and massive connectivity requirements in future cellular networks.

The rest of the paper is organized as follows. Section II explains the smart IoT and role of different computing paradigms in the realm of IoT. In section III, we present a protocol stack for smart IoT. Moreover, a detailed discussion on every layer of the stack is also provided. Finally, section IV concludes the paper.
\section{Smart Internet-of-Things}
\subsection{Data Processing on Cloud}
A cloud represents a flexible system of computational and networking nodes, usually with millions of clustered servers stacked in possibly remote physical locations, which can be accessed via the Internet. Cloud computing provides massive storage and computing platform to the organizations to store and visualize their data and that too on scalable basis i.e. pay-as-you-go fashion. The cloud allows its users to use just the right amount of resources they need and pay accordingly, while also ensuring complete flexibility to cope with unexpected bulge or drop in the demand. The services which a cloud offers are defined as X-as-a-Service: this includes the use of infrastructure such as processing and storage resources usually via virtual machines interfacing \cite{botta2016integration}. Similarly, cloud can deliver software as a service, circumventing the need for purchasing expensive licenses. Usually, the software resides on the cloud and can be used by the customers in pay-per-use fashion, making it feasible particularly for smaller businesses to manage their operational costs while also ensuring software ubiquity. Another dimension of services includes platform as a service, which provides software developers with necessary platforms in order to develop and maintain their applications without having to worry about the hardware requirements for the specific platform.

Ideally, the operations being done on the cloud can be broken into smaller fragments, which can then be assigned to different resources around the globe in a parallel fashion; this makes the cloud architecture highly robust against unexpected exigencies. However, the cloud computing does have many downsides, most important being the twofold security issues, first that there is a constant need to transfer data to and from the cloud in order to process it over the Internet and second that, since the data is stored outside an organization's infrastructure, reliance on cloud service providers increases. Other issues include increased latency of the system, which becomes a pitfall for cIoT applications where desired latency is below few milliseconds \cite{chiang2016fog}. Similarly, the data being transferred to the cloud is growing exponentially, for which restricted network bandwidth becomes an ultimate bottleneck. Moreover, most of the IoT devices such as sensors and onboard cameras have limited hardware resources and even limited network connectivity, making it impossible for these devices to send data directly to the cloud. Therefore, there is a room for some intermediary networking and computing entity which can make up for these shortcomings by drawing the computation power bit closer to the end users, while also realizing that not all kinds of data need to be cloud-sourced.
\subsection{Data Processing on Fog}
The underpinning behind the establishment of fog computing is the need for the computation, storage, and other resources to be brought closer to the end-users, complementing clouds to create a service continuum. The fog brings up the idea of distributed paradigm providing services, previously catered by the cloud, by availing all the possible resources near the end-users. It can provide massive computation power, storage, and control functions in fully distributed or centralized architectures, i.e. the applications support it provides may run anywhere between virtualized machines to dedicated hardware platforms to allow resource diversity and user mobility, ensuring common management framework for all applications.

The control functions may include applications and functions designed for end-user systems such as cyber-physical systems or services designed for cloud-based applications \cite{chiang2016fog}. This is particularly suitable for cIoT applications. The data is taken directly from the IoT sensors to the fog devices, which then process the data into necessary commands and send the control actions directly to the actuators, without having to involve the central cloud. The total resources available can be discovered using efficient resource management techniques to search for all the elements of cloud, fog, and edge in the network. Different IoT protocols can be used as bridges between edge and data centers, while software-defined networks can manage the fog networks. The fog can also assist in organizing local ad-hoc networks by integrating them with larger backbone networks and can also manage radio access networks (RAN).
\subsection{Data Processing on Edge (Device or Border Router)}
The IoT is attracting various new sectors of the society to enjoy benefits of always connected model. With the advent of several new use cases, IoT is no more a pool of dumb devices but a network of intelligent and self-sufficient beings. These intelligent beings (devices) have the capability to process data locally and to perform various tasks like data analysis and actuation, when needed. The ability to provide data analytics and processing locally on a device is termed as "edge computing". This approach can reduce computing burden and dependence on cloud/fog, while also reducing latency for cIoT applications and providing better data management using massively deployed IoT devices. Edge computing performs better when compared to its counterparts especially in cases where machine learning is required such as obstacle detection and avoidance, facial recognition system and language processing etc. More recently, multi-access edge computing (MEC) also known as mobile edge computing became part of the cellular networks in order to enable capabilities of cloud computing at the mobile edge \cite{taleb2017multi}.

Table I shows the comparison of cloud, fog, and edge computing for IoT applications against the three main phases of machine learning including; i) data pre-processing (task of preparing data according to the requirements of machine learning models), ii) model training (process of training a model using a machine learning algorithm provided a training data to learn from), and iii) model execution (testing utility and strengths of the model using a set of test data). In the case of cloud-based machine learning all three phases are centralized. For fog computing, data pre-processing and model execution are distributed while model training can be done at the cloud or cloudlet. However, in the case of full edge computing, model training is not feasible at the IoT devices as they do not have access to data from other nodes and due to their resource-constrained nature.
\begin{table}[ht]
\label{table1}
\centering
\caption{Machine Learning Phases vs different Computing Paradigm}
\begin{tabular}{|c|c|c|c|}
\hline
\textbf{\begin{tabular}[c]{@{}c@{}}Computing\\ Paradigms\end{tabular}} & \textbf{\begin{tabular}[c]{@{}c@{}}Data Pre-\\ processing\end{tabular}} & \textbf{\begin{tabular}[c]{@{}c@{}}Model\\ Training\end{tabular}}    & \textbf{\begin{tabular}[c]{@{}c@{}}Model\\ Execution\end{tabular}} \\ \hline
\textbf{Cloud Computing}                                               & Centralized                                                             & Centralized                                                          & Centralized                                                        \\ \hline
\textbf{Fog Computing}                                                 & Distributed                                                             & \begin{tabular}[c]{@{}c@{}}Centralized \&\\ Distributed\end{tabular} & Distributed                                                        \\ \hline
\textbf{Edge Computing}                                                & Distributed                                                             & Not Supported                                                        & Distributed                                                        \\ \hline
\end{tabular}
\end{table}
\subsection{Low Latency and Massive Connectivity Paradigm}
Within the regime of IoT, the system performance metrics such as latency, throughput, and energy requirements are quite different than those in traditional human-to-human (H2H) communication. In order to fully integrate different types of services into IoT and future cellular networks, data traffic emerging from machine-type communication (MTC) can be classified into two broad categories: ultra-reliable and low latency communication (URLLC) and massive MTC (mMTC). Both of these terms are defined within the realm of IoT connectivity in future cellular networks to deal with the diversity in the originating traffic.
\subsubsection{Ultra-reliable and low latency communication (URLLC)}
The URLLC, also called mission critical MTC (cMTC), defines the most critical spectrum of data traffic with stringent requirements: as evident from the name, the term is designated for the communication between nodes which demands ultra-reliability and low latency. This type of communication originates in intensive healthcare applications such as wearable sensors and remote virtual operations, in autonomous vehicles systems in which vehicles would be coordinated with each other during their course etc. All these types of applications claims extreme reliability, QoS, quality-of-experience (QoE), and non-fluctuating latency requirements as low as 1 ms as shown in Fig. 1.  For example, the connection between the communicating nodes should be dedicated at the transport layer and therefore, transmission control protocol (TCP) is the most suitable protocol owing to its reliability, while at the network layer IPv6 is preferred choice. The improvements in physical layers can be done through better channel estimation and increasing the pilot symbols to minimize the probability of error and employing efficient channel coding techniques such as, Turbo and convolutional coding \cite{ji2017ultra}. Moreover, in terms of infrastructure usage, data from MTC devices (MTDs) connected in cMTC scenarios should be intuitively shifted towards fog and edge nodes to minimize the processing time delay, since the cloud processing would possibly incur delays beyond acceptable limits.
\subsubsection{Massive machine-type communication (mMTC)}
The mMTC represents the broadest range of MTDs in IoT realm, which provide wireless connectivity to billions of devices around the globe such as wireless sensor networks in smart metering, farms, cities, and vehicles. Generally, this type of communication presents laxity in terms of delay tolerance, latency requirements, reliability, and throughput, while the major focus is on energy conservation through effective measures. This can be achieved through the use of universal datagram protocol (UDP), rather than TCP at the transport layer, which lacks feedback mechanism to guarantee reliability but saves battery life. Similarly at the network layer, 6LowPAN is more suitable due to its energy-efficient header design. Since, mMTC owns the major portion of total data traffic with higher delay tolerance, cloud processing is more appropriate for it.
\section{Protocol Stack}
In this section, we critically review the existing protocol stack for IoT in cellular networks as shown in Fig. 2, in a layerwise approach. Shaded layers in backhaul link are dependent on the technology being used at the cellular backhaul such as optical fiber cable, satellite, microwave, and free space optics links. Although, every technology has their specific protocols for backhaul communication in cellular networks, with the upcoming 5G cellular networks they pose various challenges \cite{ge20145g}.
\subsection{Data Analytics and Semantic Layer}
The volume of data generated by massively deployed IoT devices are increasing exponentially. Traditionally, this huge amount of data is stored in a data center where data processing and analytics is done. Although this approach works fine, with the introduction of 'Big Data' it has been proved inefficient. Various new schemes have been proposed recently to efficiently process big data in a device's perspective. One of the schemes is the introduction of data analytics and semantic layer at the top of the traditional protocol stack. This layer can perform data analytics locally on a device or a border router (eNodeB) using machine learning (ML) or federated learning (FL) algorithms and only segmented and related information is shared with the database/cloud.

\begin{figure}[ht]
\begin{center}
	\includegraphics[width=3in]{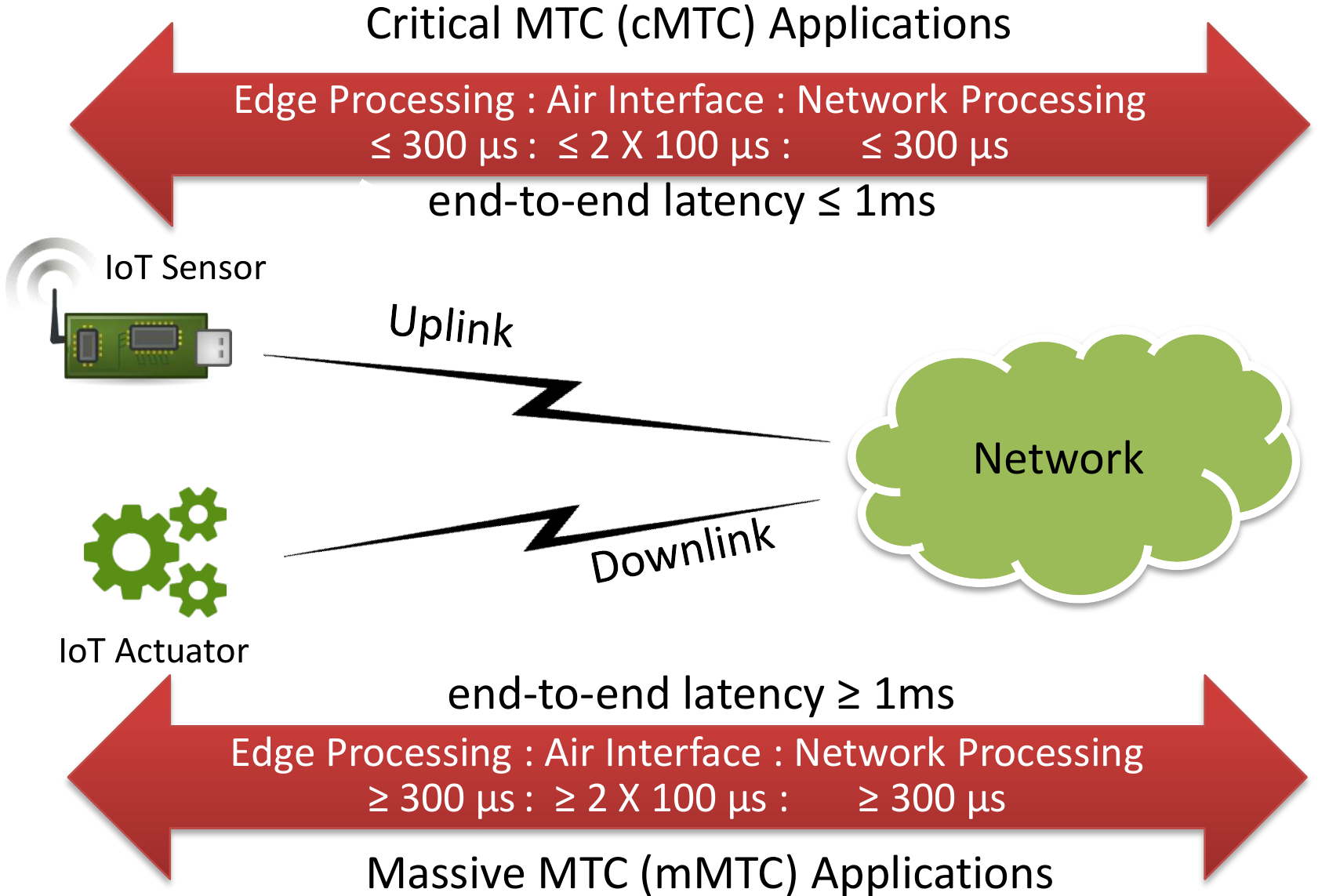}
\caption{Latency requirement for critical and massive MTC communications.}
\label{fig:ec}
\end{center}
\end{figure}
The data analytics in its true sense, contains various processes such as searching, mining, and analyzing to enhance performance. There are also various approaches such as; i) offline analytics, ii) online data analytics, iii) massive data analytics, iv) business intelligence (BI) data analytics, and v) memory level data analytics \cite{marjani2017big}. These approaches are suitable for different use cases and scenarios. The cIoT applications can switch to online data analytics which might require intelligent and resourceful (in the context of memory and energy) devices but it can speed up the process of data processing and semantic analysis. Whereas, offline data analytics is more suitable for mMTC applications due to their delay-tolerant nature. This layer can drastically reduce the workload of a data center as well as enhance network efficiency and can speed up the actuation process for various IoT applications. Moreover, efficient big data management is possible using ML, distributed ML or FL algorithms locally near the edge (device end).
\subsection{Application Layer}
Energy and resource-constrained nature of IoT devices hinder the way of conventional application layer protocols to play an important role in realizing the IoT network. The file transfer protocol (FTP) and simple mail transfer protocol (SMTP) requires high signaling overhead; thus, not suitable for most of the IoT applications. In order to address this, various new protocols have been proposed such as message queue telemetry transport (MQTT) and constrained application protocol (CoAP). MQTT was initially designed for lightweight communication of WSN and IoT. It is perfectly optimized for unreliable and lossy networks as well as high latency applications (mMTC). It follows the asynchronous mechanism with publish/subscribe architecture. MQTT can support reliable communication by ensuring three levels of QoS: i) Fire and forget; ii) Delivered at lease once; iii) Delivered exactly once \cite{itudf9kgb45}. Last but not least, it requires TCP on its transport layer which also ensures reliable communication with extra control signaling (required for acknowledgment messages).
\begin{figure}[ht]
\begin{center}
	\includegraphics[width=3in]{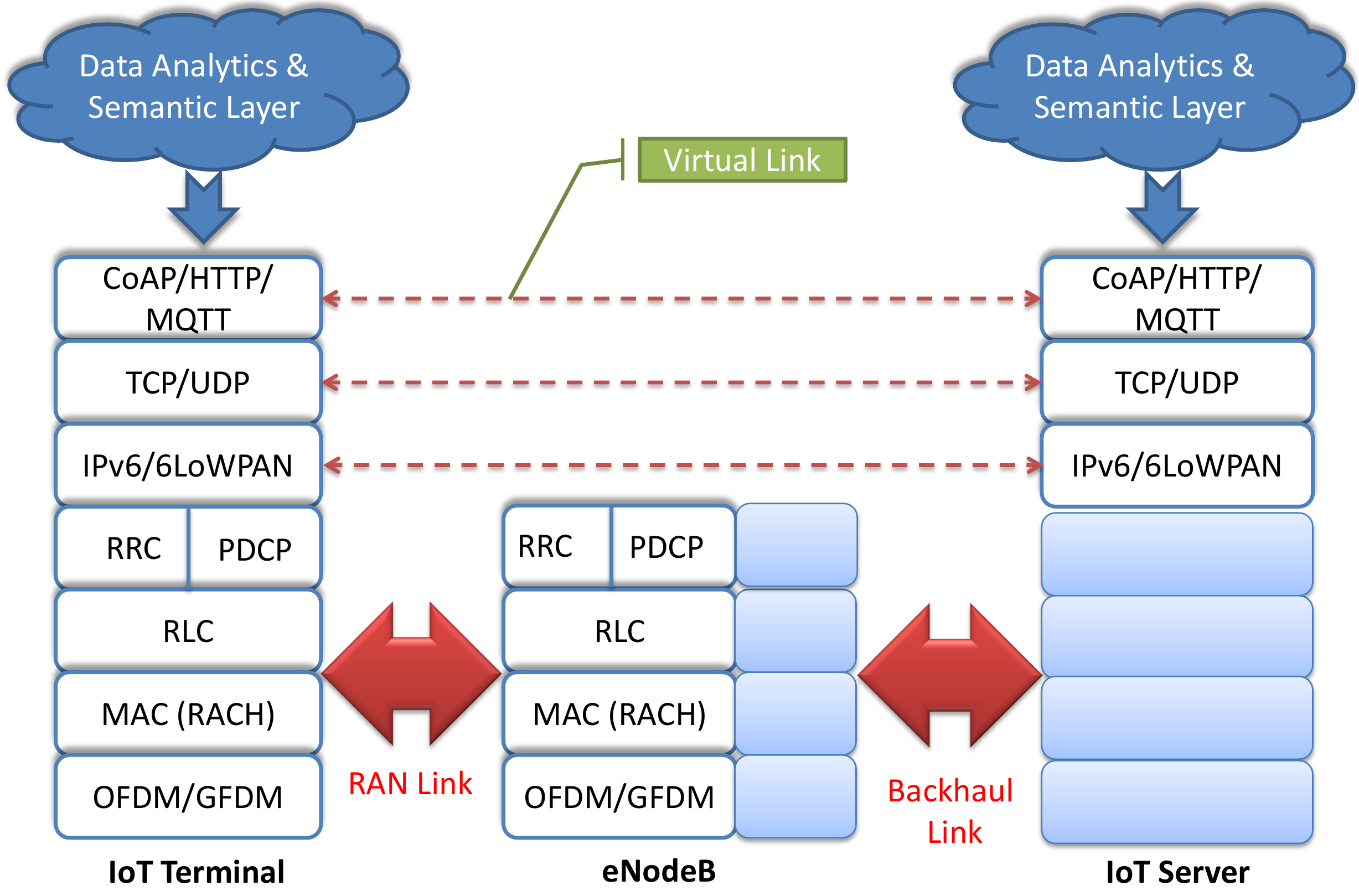}
\caption{Existing protocol stack for IoT in cellular networks.}
\label{fig:ec}
\end{center}
\end{figure}

On the other hand, CoAP was designed for synchronous communication with request/response mechanism. It requires UDP on its transport layer which does not need extra control signaling; thus, making overall implementation lightweight. Moreover, it has its own mechanism to ensure reliability through four QoS levels: i) Confirmable; ii) Non-confirmable; iii) Acknowledgment; and iv) Reset \cite{bormann2012coap}. These QoS levels help in using CoAP for various cIoT applications.
\subsection{Transport Layer}
The traditional role of the transport layer is to conduct end-to-end communication between two hosts on the Internet by establishing a logical connection between the nodes (either reliable or unreliable). The transport layer also identifies the application processes used at the source and destination nodes. Another important feature is the flow control mechanism which ensures consistent transmission between two devices with different supporting data rates.

Despite the traditionally important role of the transport layer, it cannot efficiently cope with the demands of IoT environment, mostly due to its energy-and-resource-constrained nature, as well as the low-latency requirement cIoT applications. Some research works have targeted the transport layer adaptation for the cognitive M2M networks \cite{chowdhury2009tp}. The major protocol governing the transport layer is usually TCP and occasionally UDP.

\subsubsection{TCP}
The TCP support reliable transmission of bulks of data between two applications running on different devices connected via the Internet by introducing an error-checking mechanism in its header information along with acknowledgment messages from the receiver in case of successful reception, without any support from the upper layers. Similarly, it also enables flow control to adjust the discrepancy in the transmission rates between the sender and receiver. These features make TCP a connection-oriented protocol with a focus on reliability rather than timely delivery and therefore a sub-optimal choice for IoT environment. This is further aggravated by the acknowledgment and retransmission mechanism inherited by the TCP \cite{fairhurst2002advice}. Moreover, IoT devices have high energy constraints, due to which they cannot afford large overhead associated with the header information along with the fact that most of the IoT devices are highly unlikely to maintain virtual connection due to their intrinsic sleep mode operation \cite{shang2016challenges}. Another drawback of TCP integration into IoT protocol stack is its inability to provide support for application-level framing. Application-level framing allows the breakdown of data into application data units, each of which can inherit different retransmission mechanisms to support various conditions.

\subsubsection{UDP}
In contrast to TCP, the UDP is a connectionless protocol. Although, it is less reliable than TCP, is suitable for applications where end-to-end minimal time delay is expected. Due to no acknowledgment in UDP, the chances of missed packets are high. The UDP is optimized for applications such as voice over IP (VoIP) and live video-streaming where retransmission of lost packets does not make any sense. It is for the reasons of low-latency and low power consumption that UDP is the preferred protocol for the IoT networks. The examples of recent IoT protocols which have employed UDP include BACnet and CoAP. For example, CoAP uses UDP to provide smooth implementation of IoT systems over the Internet by removing overhead of the TCP header. Moroever, CoAP also support for unicast and multicast, unlike TCP.

\subsubsection{Discussion}
IoT networks are not limited to traditional wireless connectivity, underwater and cognitive radio-based wireless sensor networks are also gaining popularity. In order to provide transport layer connectivity in these networks, various other protocols are also available. Channel-aware routing protocol (CARP) and cognitive routing protocol (CORPL) are designed for underwater wireless sensor connectivity and cognitive radio-based wireless sensor connectivity, respectively. The CARP is most suitable for resource-constrained underwater IoT connectivity due to its very small header size. Moreover, in order to provide transport layer security, datagram transport layer security (DTLS) protocol can be used. It is a stream oriented protocol which can be used in underwater wireless sensor networks for detecting eavesdroppers. The security in this protocol comes with the cost of large packet size; thus, high power consumption. 
\subsection{Network Layer}
This layer provides logical paths for data packets and therefore responsible for `data routing'. It also establishes the most optimal logical connection between source and destination and reports any failure in delivering packets. The backbone of the network layer is the Internet protocol (IP), the traditional version (IPv4) can no longer be implemented for IoT devices since it would run out of addresses when dealing with billions of these devices. The alternate version is IP version 6 (IPv6), which, however, must be tailored for low-powered IoT devices.
\subsubsection{Internet Enabled Network Layer (IPV6)}
The IPv6 is the emerging protocol for routing data packets in packet-switched networks. In order to identify the source and destinations, IPv6 uses 128-bit addresses and therefore, supports a total $2^{128}$ unique addresses. This vast array of addresses is suitable for integrating the upcoming billions of new IP based IoT devices and would also allow flexibility in address allocation. The major issue is its conformity for low power and limited bandwidth devices, which can only be made possible through IPv6 header compression.
\subsubsection{IPv6 Over Low-Power Wireless Personal Area Networks (6LoWPAN)}
This standard enables the use of IPv6 over low-power devices in order to provide Internet connectivity to even minuscule devices and therefore drastically simplifying the connectivity model for various IoT applications. The 6LoWPAN has also enabled header compression by slashing the conventional 40-byte header to allow minimum packet size of mere 4 bytes. It also eliminate the need for dynamic host configuration protocol in order to support low power networks \cite{mulligan20076lowpan}. There are multiple routing protocols have been proposed in 6LoWPAN such as lightweight ad-hoc on-demand next generation and IPv6 routing protocol for low-power and lossy networks.
\subsubsection{Discussion}
In the realm of 5G cellular networks, open cellular architecture is gaining popularity. Currently, different cellular standards including global system for mobile (GSM), enhanced data for global evolution (EDGE), universal mobile telecommunication services (UMTS), and long term evolution (LTE) are in use in different parts of the world. These standards have different network architecture and carrier requirements. In order to provide a common cellular platform for various data-centric applications, open cellular architecture based on IPv6 is gaining popularity. IPv6 can provide seamless connectivity among heterogenous networks (HetNets) which allows the introduction of innumerable new MTC applications. Through the introduction of HetNets, the problem of massive connectivity can also be resolved. Moreover, to ensure low latency and reliable communication, an evolved packet core-based architecture is suitable which employees IPv6 on its packet data network gateway (PDN-GW). This PDN-GW will enforce such policies which can provide access to cellular and non-cellular technologies altogether.
\subsection{MAC Layer}
\subsubsection{RACH Process of LTE}
\begin{figure}[ht]
\begin{center}
	\includegraphics[height=2in,width=3in]{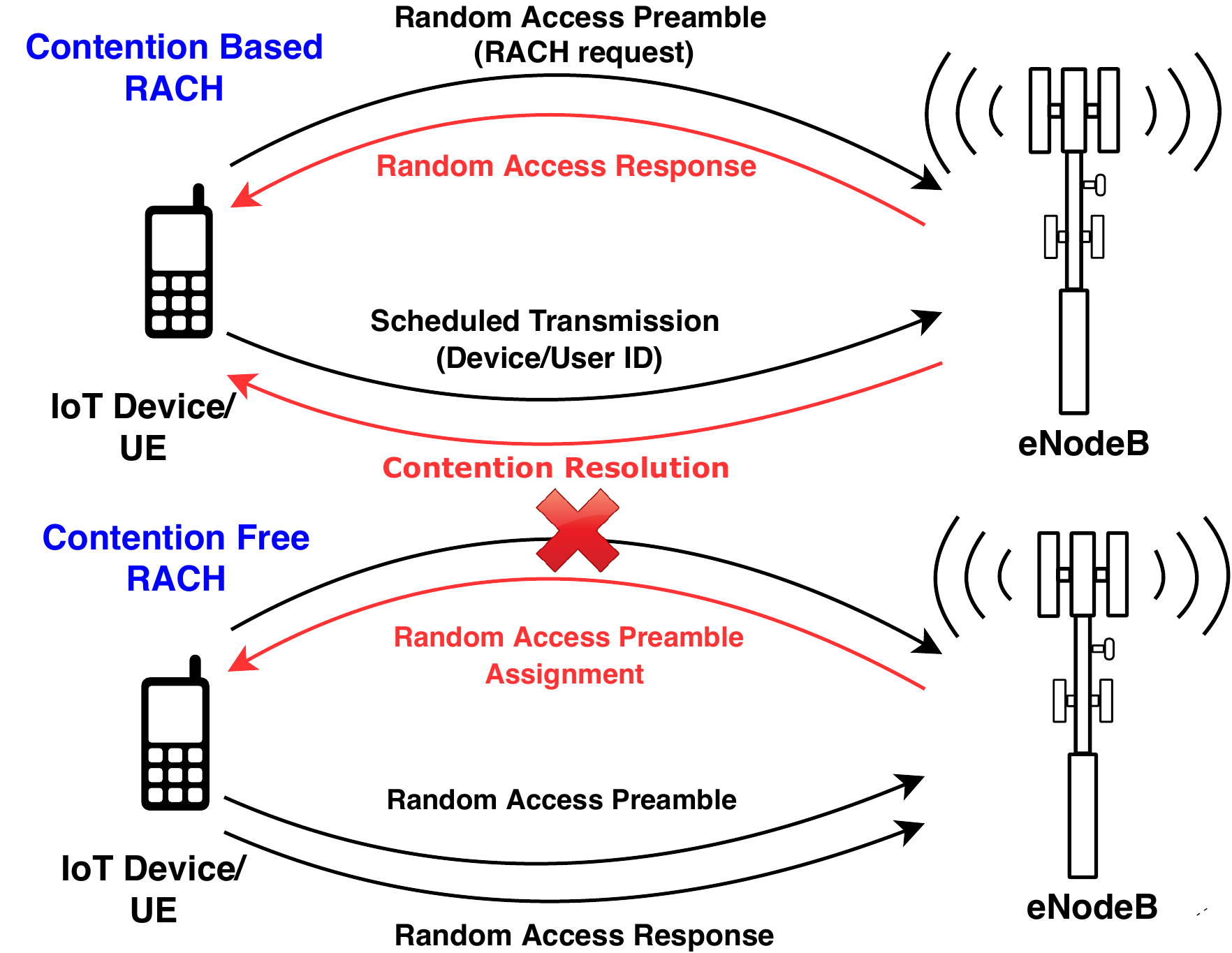}
\caption{Random access channel procedure of LTE.}
\label{fig:ec1}
\end{center}
\end{figure}

Generally in LTE, total bandwidth is divided into broadcast channels and data channels. The former require less bandwidth and are usually employed for channel access, control signaling, and broadcast purposes by eNodeB. The later occupy most of the bandwidth and are used for data communication. In LTE, user-equipments (UEs) and IoT devices contend to access data channel using physical random access channel (PRACH). The devices connected to the network can perform RACH process in contention-based, contention-free, and hybrid fashions.
In contention-based, devices present in the network initiate RACH process by randomly choosing a preamble and transmit it using the PRACH. After receiving the preamble, an eNodeB transmits random access response (RAR). If the device does not receive the RAR in a specific timing window, eNodeB retransmits the preamble after a back-off time which can be set either linearly or exponentially. This process continues until the device successfully receives the RAR. After successful reception of RAR message, the device initiates the actual RA message. In response, the eNodeB transmits a contention resolution if it receives the same preamble from multiple devices, as shown in Fig. \ref{fig:ec1}. This approach is suitable for delay-tolerant access and can also accommodate a large number of users/devices.

In contention-free, the preamble assignment is done by an eNodeB. Firstly, an eNodeB transmits the dedicated preamble information to the devices. The devices then access the channel using that specific preamble in the second step. In the end (last step), eNodeB responds with the RAR. This approach requires less control signaling for network discovery and connection; thus, more suitable for delay-sensitive access.
In order to exploit the benefits of both the approaches, a hybrid RACH mechanism can be used. It adopts the contention-based RACH when the traffic load is high and contention-free RACH when traffic load is low.
\subsubsection{Limitations of RACH Process for IoT Traffic}
Cellular networks were initially designed for H2H communication in which high QoS and reliability is one of the major key performance indicators. In recent times, cellular networks are believed to be one of the key enabling technology to envision the true landscape of the IoT network. It is expected that connected IoT devices will reach up to 50 billion by 2020 \cite{ericsson2011more}. To efficiently provide connectivity to these huge number of IoT devices, current RACH process of LTE network is not suitable. Generally in IoT, data is coming from different types of sensors comprises a few bytes (sometimes, even a couple of bits for location monitoring or weather update). This data is even smaller than the huge control signaling (required for reliable data connection in LTE). Moreover, majority of the IoT applications do not require reliable communication, thus a large control signaling overhead associated with the RACH mechanism makes it infeasible for MTC communication.

\subsubsection{Improvements in RACH Process}
First priority of cellular service providers will always be H2H users, thus separation of RA resources is one of the many proposals. In this mechanism, total RA resources are divided in two chunks, one (large) is dedicated for H2H communication and the other (small) is dedicated for MTC. Moreover, the small chunk can be further shared among MTC and H2H users; thus, allow more resources to H2H users. Generally, this approach works better due to bursty nature of MTC traffic.

Another improvement is the access class barring (ACB) mechanism, which can restrict devices from accessing the network in order to reduce the network congestion. In this mechanism, eNodeB broadcasts a random probability value to all devices available in the network. Every IoT device trying to access the network will generate a random number and if that number is smaller than the probability value, the device is allowed to attempt the channel access; contrarily, it goes into random back-off time and access is barred until the back-off timer expires \cite{shah2018congestion}. The ACB mechanism can allow an IoT device to remain in sleep mode until its back-off timer expires; thus, more energy-efficient. Besides ACB, there are various other efforts to make LTE RACH suitable for IoT data transmission which can be seen at \cite{laya2014random}.
\subsection{Physical Layer}
In this section, we present the limitations of the physical layer of LTE network and a newly proposed mechanism which claims to be an alternative to the orthogonal frequency division multiplexing (OFDM)-based physical layer. Moreover, we raised various questions on the suitability of previous and newly proposed physical layers in the context of low latency and massive connectivity.
\subsubsection{Limitations of LTE Physical Layer for Future Cellular Networks}
With the introduction of new use cases of cellular networks in various sectors of the society, the traditional physical layer structure may not fulfill the ever growing demands of cellular traffic. Although, LTE was initially designed for high throughput, low latency, and high QoS, with the advent of IoT, various new application scenarios popup which have different communication requirement when compared to H2H. These applications range from home and factory automation (high data rate and QoS required) to smart farming, and connected agriculture (very small data rate requirements, generally one message per day). Moreover, devices connected to these applications have different attributes such as, devices (sensors) used in smart farming and agriculture are battery operated; thus, energy constrained. Whereas, devices (sensors, controllers, and actuators) connected in factory automation are not energy constrained. Based on these use cases, cellular traffic is generally categorized into two types, sporadic and real-time traffic. Characteristics of a physical layer are directly dependent on these types.

Moreover, LTE network uses two basic functionalities of OFDM (synchronization and orthogonality). In Synchronization, common clock is used by the sender for processing while orthogonality ensures that no crosstalk occurs in the receiver's waveform detection process. These functionalities are power and bandwidth hungry; thus, to accommodate the huge number of IoT devices new approaches are the need of a day.
In order to address the shortcomings of existing physical layer of LTE network, fifth-generation non-orthogonal waveform (5GNOW) project has proposed the use of generalized frequency division multiplexing (GFDM) instead of OFDM \cite{wunder20145gnow}. GFDM is a multi-carrier system which employs cyclic prefix at the transmitter end for low complex equalizer. It can exploit spectrum fragmentation by employing non-orthogonal waveforms. It also supports multi-user scheduling in both time and frequency domain. For IoT connectivity, it exhibits better performance when compared to OFDM due to its non-orthogonal nature, low power consumption and reduced hardware cost \cite{fettweis2009gfdm}.
\subsubsection{Discussion}
Current physical layer approaches are not feasible for fulfilling the basic requirements of MTC communication in future cellular networks. There arise various questions such as: Should orthogonality be a part of channel access? Does the unified uplink frame structure ensure massive connectivity? Does reliability deteriorate by using non-orthogonal waveforms? How to ensure ultra-low power communication for MTC device? In order to address these questions effectively in upcoming 5G cellular networks, new techniques such as, compressive-sensing, coded random access, digital signal modulation through deep-learning, and convolutional neural network-based modulation classification must be incorporated. Moreover, there are several emerging technologies for IoT connectivity such as machine-to-machine, device-to-device, and vehicle-to-vehicle communications and in order to deploy a reliable end-to-end system, it is critical to understand their physical layer characteristics and requirements.

Moro recently, IEEE 1932.1 working group has started working on a standard for interoperability among devices designed for licensed and unlicensed frequency spectrum, including Wi-Fi and LTE. This standard will provide a mechanism for interoperation of MAC and physical layer protocols designed for various licensed and unlicensed spectrum \cite{mumtaz2018guest}. Physical layer interoperability for IoT devices will bring in many challenges including spectrum selection, power consumption during spectrum sensing/scanning and increase in hardware module cost. 
\section{Conclusion}
In this work, we discussed different storage and computing models to account for the tremendous growth of data that would be generated as number of IoT devices are growing exponentially. We have identified various issues in IoT protocols for different layers which may bound low latency and massive connectivity requirements in future cellular networks. Most frequently used protocols at application layer (MQTT and CoAP) may provide massive connectivity in mMTC applications with high latency and high reliability in cMTC applications with low connectivity. At the transport layer, TCP may provide high reliability and low latency with more power consumption when compared to UDP. At the network layer, 6LowPAN can provide access to resource-constrained IoT devices but due to the data-centric nature of new cMTC applications, need for a common basis using IPv6 for future cellular networks is gaining popularity. Extending the discussion to MAC and physical layer, we raised several questions related to orthogonal nature of channel access and does newly propose non-orthogonal solutions provide reliability.

\section*{Acknowledgment}
This work is partially supported by the National Natural Science Foundation of China under Grant number 61871190.
\appendices

\footnotesize{
\bibliographystyle{IEEEtran}
\bibliography{references}
}

\vfill\break
\end{document}